\documentclass[final,5p,times,twocolumn,sort&compress]{elsarticle}

\usepackage{graphicx}

\usepackage{amsmath}
\usepackage{amssymb}
\usepackage{textcomp}

\usepackage{wasysym}

\usepackage{lineno}

\usepackage{epstopdf}
\usepackage{natbib} 

\usepackage{hyperref}  
\hypersetup{colorlinks = true}

\journal{Phys. Lett. B}

\begin{document}

\begin{frontmatter}

\title{Measurement of the $p(e,e'\pi^+)n$ reaction close to threshold and at low $Q^2$}

\author[pmf]{I. Fri\v{s}\v{c}i\'{c}\corref{cor1}\fnref{fn1}}
\author[kph]{P.~Achenbach} 
\author[kph]{C.~Ayerbe Gayoso} 
\author[kph]{D.~Baumann} 
\author[kph]{R.~B\"{o}hm} 
\author[pmf]{D.~Bosnar}   
\author[jsi]{L.~Debenjak} 
\author[kph]{A.~Denig} 
\author[kph]{M.~Ding} 
\author[kph]{M.~O.~Distler}
\author[kph]{A.~Esser} 
\author[kph]{H.~Merkel} 
\author[kph]{D.~G.~Middleton}  
\author[kph,jsi]{M.~Mihovilovi\v{c}}  
\author[kph]{U.~M\"{u}ller}  
\author[kph]{J.~Pochodzalla}  
\author[kph]{B.~S.~Schlimme}  
\author[kph]{M.~Schoth}   
\author[kph]{F.~Schulz}  
\author[kph]{C.~Sfienti}  
\author[dplj,jsi]{S.~\v{S}irca} 
\author[kph]{M.~Thiel} 
\author[kph]{Th.~Walcher} 

\cortext[cor1]{Corresponding author, Email: friscic@mit.edu}
\fntext[fn1]{Present address: MIT-LNS, Cambridge MA, 02139, USA}

\address[pmf]{Department of Physics, Faculty of Science, University of Zagreb, HR-10002 Zagreb, Croatia}
\address[kph]{Institut f\"{u}r Kernphysik, Johannes Gutenberg-Universit\"{a}t, D-55099 Mainz, Germany}
\address[jsi]{Jo\v{z}ef Stefan Institute, SI-1000 Ljubljana, Slovenia}
\address[dplj]{Department of Physics, University of Ljubljana, SI-1000 Ljubljana, Slovenia}

\begin{abstract}
The cross section of the $p(e,e'\pi^+)n$ reaction has been measured for 
five kinematic settings at an invariant mass of $W = 1094$ MeV and for a 
four-momentum transfer of $Q^2 = 0.078$ (GeV/$c$)$^2$. The measurement 
has been performed at MAMI using a new short-orbit spectrometer (SOS) of 
the A1 collaboration, intended for detection of low-energy pions. The 
transverse and longitudinal cross section terms were separated using the 
Rosenbluth method and the transverse-longitudinal interference term has 
been determined from the left-right asymmetry. The experimental cross 
section terms are compared with the calculations of three models: DMT2001, 
MAID2007 and $\chi$MAID. The results show that we do not yet understand 
the dynamics of the fundamental pion.
\end{abstract}


\end{frontmatter}

\section{Introduction} \label{introduction}
\begin{figure}
\includegraphics[height=340pt, width=160pt]{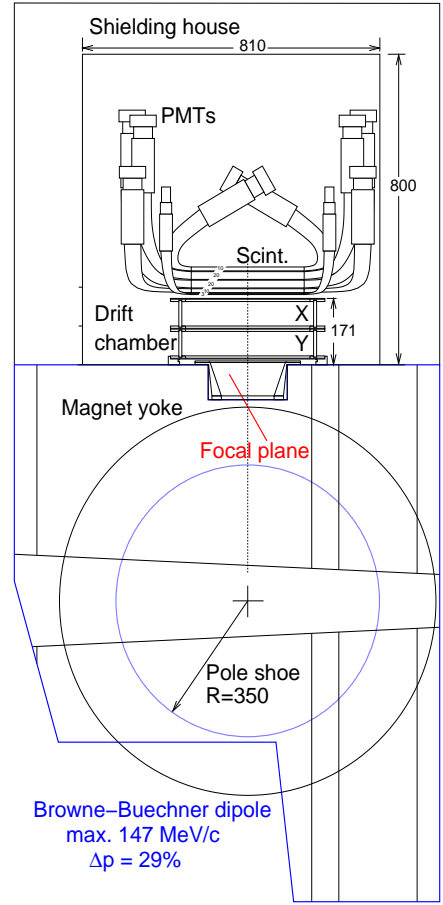} \centering
\caption{(colour online) The SOS spectrometer with the detector system. The measures are 
in mm, the target position is 1131.6 mm to the right in line with the central 
cross. \label{fig_SOS}}
\end{figure}

Pion electroproduction on protons near threshold has been
in the focus of both theoretical and experimental research for more than 
four decades. The pioneering theoretical studies were performed by deriving 
low-energy theorems (LETs) based on chirality conservation for soft pions, 
which established the connection between the S-wave transverse multipole of 
the charged pion electroproduction and the axial form factor 
\cite{Nambu1962_1, Nambu1962_2}. The same result was later derived in the 
framework of current algebra using the partially conserved axial current 
(PCAC) and, additionally, a relation between the S-wave longitudinal
multipole and the induced pseudoscalar form factor was derived 
\cite{Amladi1979}. After the formulation of chiral perturbation theory 
($\chi$PT) new calculations showed \cite{Bernard1992, Bernard1993} that 
the pion loop contributions to the LETs could not be neglected as it was the 
case in previous approaches \cite{Vainshtein1972, Scherer1991}.
\newline \indent Compared to the comprehensive experimental data for pion
photoproduction \cite{Arndt_2006, Ronchen_2013, Drechsel_2007_47} (and references therein), the 
existing data for electroproduction, i.e., at non-vanishing four-momentum 
transfer, are sparse. This is especially true for the charged pion channel 
close to threshold, where detection of either a recoiling neutron or a pion 
at low energy makes a measurement difficult.
The earliest charged pion electroproduction experiments were carried out in the 1970s, 
but their results had large statistical and systematic uncertainties. 
They can be divided into single arm experiments 
as in Stanford (SLAC) \cite{Bloom1973} and Kharkov \cite{Esaulov1978}, or 
double arm coincidence electron and neutron experiments as in Frascati 
\cite{Amladi1972}, Hamburg (DESY) \cite{Brauel1973, Joos1976} and Daresbury 
\cite{delGuerra1975, delGuerra1976}. From the 1990s and onwards, electron and 
pion coincidence experiments were performed in Saclay \cite{Choi1993}, JLab 
(CLAS) \cite{Park2012} and in Mainz (MAMI) \cite{Blomqvist1996, Liesenfeld1999}. 
\newline \indent In the one-photon-exchange 
approximation, the differential cross section of the $p(e,e'\pi^+)n$ reaction 
can be expressed as a product of the virtual photon flux $\Gamma$ and the 
virtual photon cross section $d\sigma_\nu/d\Omega^\star_\pi$ 
\cite{Amladi1979, Drechsel1992}. For an unpolarised electron beam and an 
unpolarised target, $d\sigma_\nu/d\Omega^\star_\pi$ can be further factorized 
into one transverse ($T$), one longitudinal ($L$) and two interference ($TL$ 
and $TT$) terms (convention from \cite{Drechsel_2007_47}):
\begin{equation}
  \begin{aligned} \label{eq:cross} 
\frac{d\sigma_\nu}{d\Omega^\star_{\pi}} &= \frac{d\sigma_T}{d\Omega^\star_\pi}\! + \!
\epsilon \frac{d\sigma_L}{d\Omega^\star_\pi}\! +\!\! \sqrt{2\epsilon (1 + \epsilon)}
\frac{d\sigma_{T\hspace{0.1mm} L}}{d\Omega^\star_\pi}\!\cos\phi_\pi\! + \!\epsilon \frac{d\sigma_{T\hspace{0.1mm} T}}{d\Omega^\star_\pi}\!\cos2\phi_\pi \\
\epsilon &= \Bigg(1+ \frac{2 \vec{q} ^2}{Q^2} \tan^2\frac{\theta_{e^-}}{2} \Bigg)^{-1}  
\end{aligned}
\end{equation}
\noindent where $\phi_\pi = \phi^\star_\pi$ is the angle between the scattering and 
the reaction plane, the asterisk denotes quantities evaluated in the hadronic 
centre-of-mass frame. $\epsilon$ is the transverse polarization of the virtual 
photon and at fixed $Q^2$ the value of $\epsilon$ can be selected by using
the right combination of beam energy and electron scattering angle $\theta_{e^-}$.
The $T$ and the $L$ terms can be separated by a measurement in the 
so-called parallel kinematics (the pion production angle 
$\theta_\pi = \theta^{\star}_\pi = 0^\circ$). There, the interference terms vanish due 
to their angular dependence, $d\sigma_{T\hspace{0.1mm} L} \sim 
sin \theta^{\star}_\pi$ and $d\sigma_{T\hspace{0.1mm} T} \sim sin^2 \theta^{\star}_\pi$ 
\cite{Amladi1979, Drechsel1992}. The virtual photon cross section \linebreak 
is determined for fixed values of the invariant mass $W$ and four-momentum 
transfer $Q^2$, by varying only $\epsilon$. The separation of $T$ and $L$ 
terms is then performed using the Rosenbluth method \cite{Rosenbluth1950}. 
The determination of the $TL$ term requires two measurements at fixed values 
of $W$, $Q^2$, $\epsilon$ and $\theta^{\star}_\pi \neq 0$. One measurement 
is performed at $\phi_\pi = 0^\circ$ and the other at $\phi_\pi = 180^\circ$ 
\cite{Drechsel1992}. The $TL$ term can then be determined from the left-right 
asymmetry:
\begin{equation} \label{eq:TL1}
\frac{d\sigma_{T\hspace{0.1mm} L}}{d\Omega^\star_\pi} = \dfrac{\dfrac{d\sigma}
{d\Omega^\star_\pi}\Big|_{\phi_\pi = 0^{\circ}, \theta^\star_\pi \ne 0^{\circ}} 
- \dfrac{d\sigma}{d\Omega^\star_\pi}\Big|_{\phi_\pi = 180^{\circ}, 
\theta^\star_\pi \ne 0^{\circ}}}{2\sqrt{2\epsilon(1 + \epsilon)}}
\end{equation} 
Knowing the precise values of the experimental cross section terms $T$, $L$ 
and $TL$ at a given $W$ and $Q^2$ allows the testing of the theoretical 
predictions of these quantities, which in turn are based on various approaches 
in describing the structure of the nucleon.  
\newline \indent Experiments in Mainz were performed at 46 MeV above the production threshold, 
using the high precision spectrometer setup of the A1 Collaboration 
\cite{Blom1998}. Since the particle path in these spectrometers is of the order of 10 m, 
in an experiment closer to threshold most pions would decay before reaching
detectors. This would lead to contamination of the data with
muons indistinguishable from pions, thus increasing systematic
errors beyond acceptable limits. A spectrometer with a shorter path 
significantly reduces the muon contamination.  
\newline \indent In this Letter we present the measurement and analysis of a 
$p(e, e'\pi^+)n$ coincidence experiment at only 15 MeV above the threshold. 
The experiment was performed at a fixed invariant mass of $W$ = 1094 MeV and 
virtual photon four-momentum transfer of $Q^2$ = 0.078 (GeV/$c$)$^2$. 
The low-energy pions were detected in a new short-orbit spectrometer (SOS, see 
Fig \ref{fig_SOS}) \cite{Baumann2015} with a particle path of $\approx$ 1.6 m. 
The obtained $T$, $L$ and $TL$ terms were compared with predictions of three 
different models.  

\section{Experiment} \label{experiment}
The experiment was carried out at the 
spectrometer setup of the A1 Collaboration \cite{Blom1998} at the Mainz 
Microtron \cite{Herminghaus1976}. The energy of the unpolarised electron beam 
was varied between 345 and 855 MeV. The beam current ranging from 7 to 25 
$\mu$A  was measured using a  flux-gate magnetometer. A cylindrical target 
cell with a diameter of 2 cm and Havar walls of 50 $\mu$m was used in 
combination with a high power liquid hydrogen cooling system. In order to 
avoid density fluctuations, the beam was rastered in transversal directions 
and the liquid hydrogen was recirculated. 
\newline \indent The scattered electron was detected in the standard spectrometer A, 
while the produced charged pion was measured with the SOS. 
In the electron arm, four vertical drift chambers were used for particle 
tracking, while in the pion arm the tracking detector was realized with two 
volume type drift chambers using a helium-ethane mixture as counting gas. In 
both arms, scintillation detectors were used for trigger and timing purposes, 
and additionally, they were used for particle identification in the pion arm. 
The angular acceptances of the spectrometers were defined by heavy metal 
collimators: 21 msr for spectrometer A and 2 msr for the SOS. The momentum 
acceptance was 20\% and 29\% in spectrometer A and SOS, respectively.
\newline \indent The kinematic settings are summarized in Table \ref{table1}. 
The settings 1, 2 and 3 were measured in parallel kinematics for the 
separation of $T$ and $L$ terms. The last two settings were measured at 
$\theta^{\star}_{\pi} =\pm$ $18.7^{\circ}$ with respect to the virtual photon 
direction, to determine the $TL$ term.
\begin{table}[!ht]
\caption{Central values of the kinematic settings: $E$ is the energy of the 
beam, $p_{e^-}$ is the momentum of the scattered electron, $\theta_{e^-}$ is 
the angle of the scattered electron, $p_{\pi^+}$ is the momentum of the pion 
and $\theta_{\pi^+}$ is the pion production angle.}
\setlength\tabcolsep{4pt}  
\begin{tabular}{ccccccc} 
Setting &$\epsilon$& $E$ &$p_{e^-}$&$\theta_{e^-}$&$p_{\pi^+}$&$\theta_{\pi^+}$\\
        &      &(MeV)&(MeV/$c$)&($^{\circ}$)  & (MeV/$c$) &  ($^{\circ}$)  \\ \hline
1       &0.3065& 345 & 134.8   & 80.7         &    113    & 22.40          \\ 
2       &0.5913& 450 & 239.8   & 50.3         &    113    & 31.79          \\ 
3       &0.8970& 855 & 644.8   & 22.5         &    113    & 42.94          \\ 
4       &0.8970& 855 & 644.8   & 22.5         &    110    & 32.80          \\
5       &0.8970& 855 & 644.8   & 22.5         &    110    & 53.10          \\
\end{tabular} 
\label{table1} 
\end{table}

\section{Data analysis}  \label{dataanalysis}
\begin{figure} 
\includegraphics[height=185pt, width=240pt]{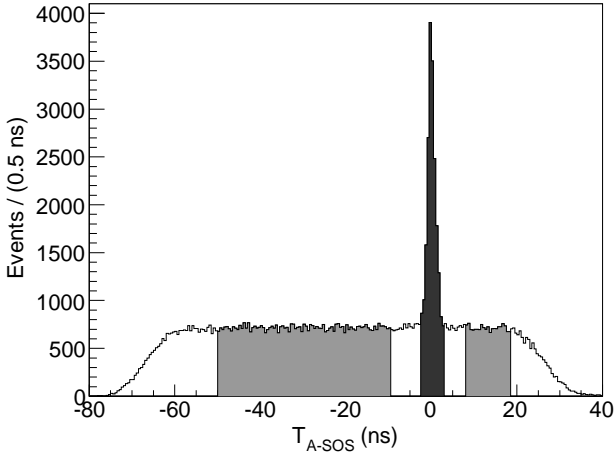} \centering
\caption{\label{fig:time} Coincidence time distribution. The dark gray area includes 
the true coincidences and the light gray area contains only random coincidence events 
used for the background estimation.}
\end{figure}

\begin{figure}[!hb]
\includegraphics[height=187pt, width=240pt]{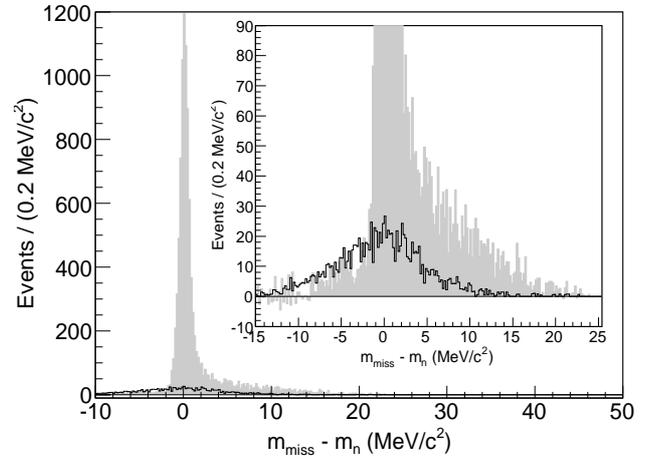}  \centering
\caption{\label{fig:mm} The background subtracted missing mass distribution 
of setting 2 relative to the neutron mass (light gray). The insert shows the plot in the peak region at 
a magnified scale. The black line denotes the muon distribution from the simulation of the pion decay.}
\end{figure}

The true events were first identified by 
measuring the coincidence time between spectrometer A and SOS, which was 
corrected for the path length of the particle in the corresponding spectrometer 
and for delays in the electronics. After these corrections, a sharp 
coincidence peak of 2.8 ns FWHM was obtained, see Fig. \ref{fig:time}. A cut at 
-2 ns $\le$ $\Delta$t $\le$ 2.5 ns was used on the coincidence peak to select the true 
electron-pion pairs. Cuts on sidebands at -50 ns $\le$ $\Delta$t $\le$ -10 ns 
and 7 ns $\le$ $\Delta$t $\le$ 18 ns were used to estimate the background 
contribution from random coincidences. In the case of spectrometer A,
further reduction of the background was achieved by cuts on the reconstructed electron 
momentum and the reconstructed vertex. For the SOS, this included cuts on 
acceptance of the dispersive angle and reconstructed pion momenta. A special 
cut and correction was applied for particles passing very close to the signal 
wires of the SOS drift chamber \cite{Baumann2015} which affected approx. 2\% 
of the events.
\newline \indent The measured electron and pion momenta were used to calculate 
the missing mass of an unobserved neutron for each event. Additionally, 
particle momentum was corrected for energy loss in the target and in different 
materials. The random coincidence background was removed by subtraction. As an
example, the resulting missing mass distribution of setting 2 is
shown in Fig. \ref{fig:mm}. Events between -3 and 11 MeV/$c^2$ relative to the neutron mass 
were selected as the final true events. These events were corrected for 
detector inefficiencies (varying from 16.2\% to 21.0\%) as well as for 
in-flight pion decay. The calculation of the  decay correction factor was 
based on the determined pion trajectory length from the interaction vertex 
to the scintillation detectors and the measured pion momentum. 
\newline \indent Integrated luminosity was calculated off-line and corrected 
for dead time (varying from 5.4\% to 7.8\%). The target density value was 
updated if the temperature changed by more than 0.03 K and the pressure by more 
than 2 mbar, whereas smaller fluctuations were averaged out.
\begin{figure}[!hb]
\includegraphics[height=187pt, width=240pt]{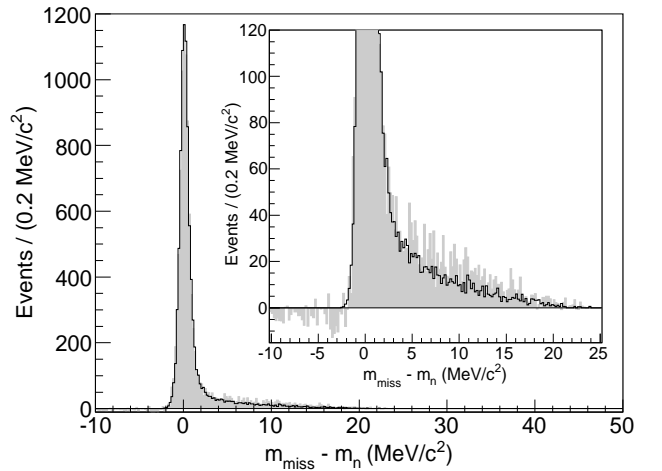}  \centering
\caption{\label{fig:mm3} Missing mass distribution of setting 2 (light gray) 
with the muon distribution subtracted. The insert shows the radiative tail region 
at a magnified scale. The black line represents the distribution of the accepted 
phase space for setting 2. Experimental and simulated were made comparable
by requesting that the highest bins of both distributions have the same value.}
\end{figure}
\newline \indent The phase space accepted by the apparatus was determined based on a Monte 
Carlo simulation, which provided event by event corrections stemming from 
radiative (internal and external Bremsstrahlung, vertex correction) and ionization 
losses, and also included the angular and momentum 
resolution parameters of the spectrometers. Muons originating from in-flight 
pion decays, created near the detectors or close to the direction of the 
decayed pions, were detected and a certain amount of these could not be 
distinguished from pions. Therefore, a Monte Carlo simulation of pion decays 
inside the SOS was performed. The simulation incorporated full tracking of 
a pion or produced muon trajectory from the vertex, through the simulated 
magnetic field of the SOS dipole magnet, to the scintillation detectors. 
In order to make distributions form the experiment and simulation comparable, 
the muon distribution was normalized with the ratio of the highest bins of the 
experimental distribution and the combined simulated pion+muon distribution. Fig. 
\ref{fig:mm} (insert) shows the muon missing mass distribution (denoted by a 
black line). As expected, muons under the true events distribution (shaded light gray) 
will be misidentified as pions. Furthermore, the shape of the muon distribution 
follows the "foot" at the left edge of the true events distribution. 
\newline \indent Naive subtraction of the muon distribution removes this structure 
and assures that the left edge of the experimental distribution agrees with the left 
edge of the phase space distribution (see Fig. \ref{fig:mm3}). This confirms that the "foot" 
is caused by muons. Due to low statistics of the experimental data in the "foot" region, 
the subtraction introduces unphysical negative values in the experimental distribution. 
To avoid this when calculating the cross section, but also to prevent possible normalization 
issues, muons are not removed from experimental data using subtraction. Instead the muon 
percentage inside a certain missing mass interval is calculated based on simulation. 
For the missing mass interval from -3 to 11 MeV/$c^2$ a muon contamination varying 
from 7.3\% to 7.8\% was obtained.
\begin{table}
\caption{Individual systematic uncertainties.} 
\setlength\tabcolsep{4pt} 
\begin{tabular}{lc}
Source                 & Relative uncertainty (\%) \\ \hline
Layer of residual gases&             0.10          \\
Phase space            &             0.12          \\
Muon contamination     &             0.23          \\
Pion decay correction  &             0.54          \\
Luminosity             &             0.58 - 0.59   \\
Badly reconstructed tracks (SOS) &   0.68          \\ 
Missing mass cut       &             2.44 - 3.20   \\ 
All other cuts         &             0.96          \\ \hline
Total (quadrature sum) &             2.9 - 3.5     \\
\end{tabular} 
\label{table2}
\end{table} 
\newline \indent Table \ref{table2} contains individual and total relative 
systematic uncertainties. Given that we used a cryogenic target, a layer of 
residual gases formed on the target surface due to the non-perfect vacuum. 
The contribution of this layer was estimated by changing its thickness 
in the simulation. Contributions due to phase space and muon contamination 
were determined by simulation. The uncertainty of the pion decay correction 
was contributed by the errors of trajectory length and reconstructed pion 
momentum. Luminosity uncertainty comprises errors stemming from target density 
fluctuations, measurement of beam current and determination of the average 
target length. The fraction of corrected events per drift cell was used to 
estimate the contribution from badly reconstructed tracks. The main systematic 
uncertainty is due to the cut-off on the radiative tail of the missing mass 
distribution. This contribution depends on corrections for radiative processes which were applied
in the simulations, the subtraction of the random background and the fluctuation of
the data due to low statistics in the radiative tail (see Fig. \ref{fig:mm3} insert).

\section{Results and discussion}  \label{results}

\begin{figure*}[!ht]
\includegraphics[height=308pt, width=510pt]{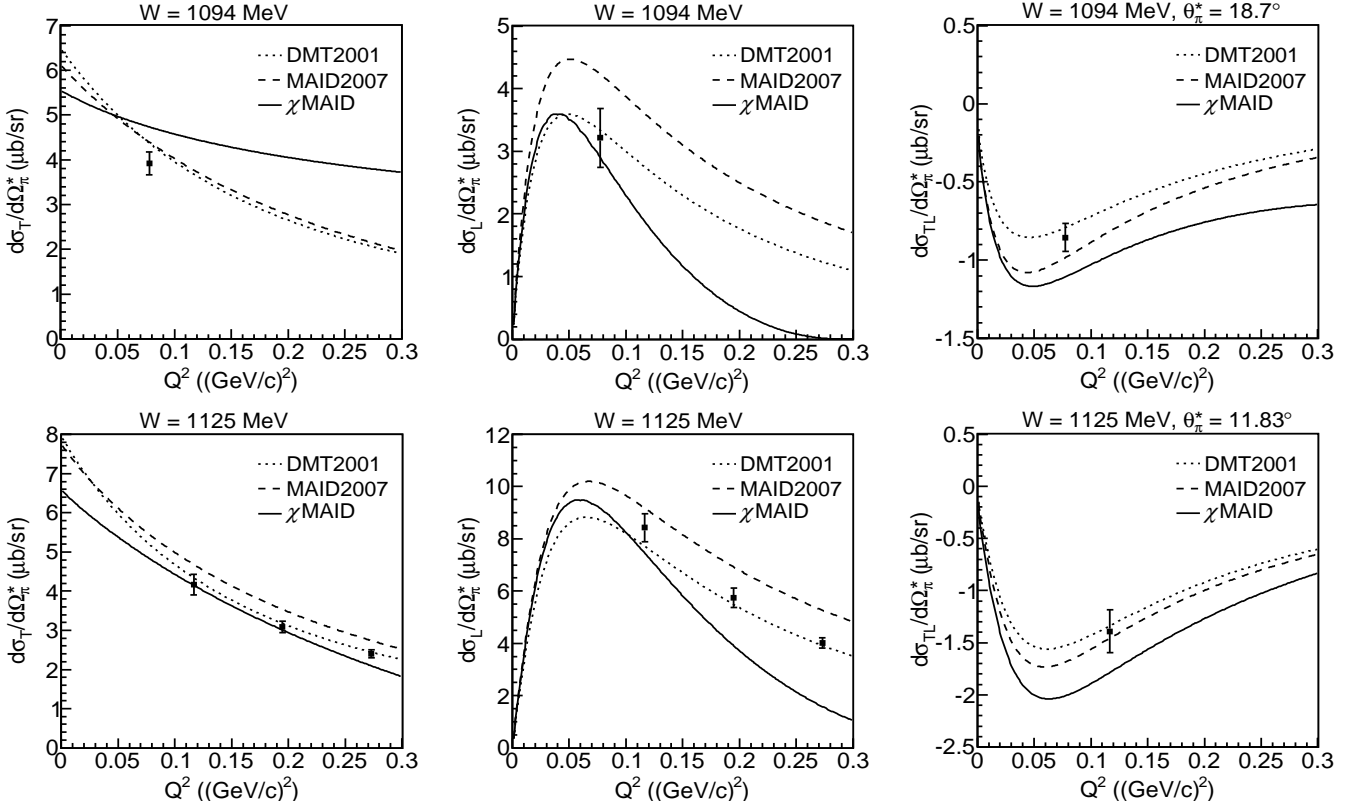} \centering
\caption{\label{fig:exp_and_the} Experimental T, L and TL terms with the total
errors plotted together with model predictions as a function of Q$^2$. 
The figures in the first row show results at W = 1094 MeV, reported in this Letter.
The second row shows previously published results from experiments at
MAMI, measured at W = 1125 MeV \cite{Blomqvist1996, Liesenfeld1999}.}
\end{figure*}

Measured cross sections of the $p(e,e'\pi^+)n$ reaction, with the statistical, systematic 
and total errors, are listed in Table \ref{table3}. 
The $L$ and $T$ terms (see Table \ref{table4}) were extracted by performing 
a weighted linear regression on 
settings 1, 2, and 3, using the total errors as weights.
The $TL$ term and corresponding total 
error were determined from the left-right asymmetry and Eq. (\ref{eq:TL1}).
\begin{table}
\caption{Measured $p(e,e'\pi^+)n$ cross sections.} 
\setlength\tabcolsep{3.5pt} 
\normalsize
\begin{tabular}{cccccc}
Setting&$d\sigma/d\Omega^\star_{\pi}$&$\pm$&Total Error  &Stat. Error  &Syst. Error \\  
       &    ($\mu$b/sr)              &     &($\mu$b/sr)  &($\mu$b/sr)  &($\mu$b/sr) \\ \hline 
1      &  4.91                       &$\pm$& 0.15 (3.1\%)&0.06 (1.2\%) &0.14 (2.9\%)\\  
2      &  5.73                       &$\pm$& 0.21 (3.7\%)&0.06 (1.1\%) &0.20 (3.5\%)\\   
3      &  6.83                       &$\pm$& 0.24 (3.5\%)&0.08 (1.2\%) &0.23 (3.4\%)\\  
4      &  5.37                       &$\pm$& 0.18 (3.4\%)&0.05 (0.9\%) &0.17 (3.1\%)\\  
5      &  8.53                       &$\pm$& 0.27 (3.2\%)&0.07 (0.8\%) &0.26 (3.1\%)\\   
\end{tabular}
\label{table3}
\end{table}
\begin{table}
\caption{Experimental and model results for the $L$, $T$ and $TL$ terms of 
the charged pion electroproduction on protons at $W$ = 1094 MeV and at 
$Q^2$ = 0.078 (GeV/$c$)$^2$.} 
\setlength\tabcolsep{2.5pt} 
\begin{tabular}{ccccc}
Term& Data $\pm $ Total Error & DMT2001   & MAID2007  & $\chi$MAID \\  
    &     ($\mu$b/sr)         &($\mu$b/sr)&($\mu$b/sr)&($\mu$b/sr) \\ \hline 
$T$ &3.91  $\pm$ 0.26 (6.7\%) &  4.38     & 4.39      & 4.73       \\  
$L$ &3.20  $\pm$ 0.47 (14.7\%)&  3.33     & 4.22      & 2.89       \\   
$TL$&-0.86 $\pm$ 0.09 (10.5\%)& -0.79     & -0.98     & -1.11      \\  
\end{tabular}
\label{table4}
\end{table}
\newline \indent The experimental terms are compared with predictions given 
by three state-of-the-art theoretical models \cite{web_models_43}. The first 
model is the Dubna-Mainz-Taipei (DMT2001). The calculations were done within 
a meson-exchange dynamical model, which uses potentials derived from an 
effective chiral Lagrangian \cite{Kamalov_44, Kamalov_revC_45, Kamalov_revLett_46, Yang_2012, Yang_2016}. 
The second model is based on a partial-wave analysis using the Mainz unitary 
isobar model MAID2007 \cite{Drechsel_2007_47}. The third model is the Chiral 
MAID ($\chi$MAID) \cite{Hilt_2013_49, Hilt_2016}. It calculates the pion photo- and 
electroproduction based on Lorentz-invariant baryon chiral perturbation theory 
up to and including order $q^4$. The low-energy constants (LECs) are fixed by 
fitting the previous experimental data in all available reaction channels. For 
this Letter, the LECs were not changed.   
\newline \indent The obtained experimental results and model predictions for 
the $L$, $T$ and $TL$ cross section terms are presented in Table \ref{table4}. 
The errors are the so called uncorrelated errors, better called one-parameter 
errors (see Chap. 9 of ref. \cite{James:2006zz}). The deviation of the 
theoretical values from the experimental results is of the order of one to 
three standard deviations of these errors customarily taken. However, these 
errors feign an agreement which does not exist. 
If one includes the correlation between the $T$ and the $L$ terms one can 
construct a 3d-error ellipsoid. The distance between the experimental values 
and the model predictions, using the principle axes of the error ellipsoid as 
a coordinate system, is a measure of the simultaneous agreement between the 
theoretical terms and the experiment. One gets a \linebreak p-value of $2 \times 10^{-5}$ 
for the DMT2001 model, $6 \times 10^{-10}$ for $\chi$MAID, and $ < 10^{-16}$ 
for MAID2007. The DMT2001 model has some edge over the others. 
It is based on a realistic pion-nucleon interaction and includes chiral symmetry 
from the start. Compared to $\chi$PT, the dynamic approach takes all loops up 
to the arbitrary order into account \cite{Kamalov_revLett_46}. This makes it 
superior with respect to any $\chi$PT calculation. On the other hand, the 
predictions of $\chi$MAID depend strongly on LECs which are fixed by only few 
experimental results \cite{Hilt_2013_49}. Adding of our results would lead to 
estimates of different LEC values, which may in turn better reproduce our 
$T$, $L$ and $TL$ terms.
\newline \indent Fig. \ref{fig:exp_and_the} shows plots of models and the new 
results reported here, measured at W = 1094 MeV, and the previous MAMI 
measurements at W = 1125 MeV \cite{Blomqvist1996, Liesenfeld1999}.
Even for the previous MAMI results, DMT2001 is overall in better agreement with
the experiment, compared to the other two models. The same was also 
concluded from our terms. 
$\chi$MAID describes very well the T term data points of the previous MAMI results, better than MAID2007, but this is not a surprise since the used LECs
were obtained by fitting the W = 1125 MeV data \cite{Hilt_2013_49}. This situation is reversed for our data.   
The TL terms  are better described by MAID2007  than by $\chi$MAID. Both models have difficulties in describing the L term data points of the previous MAMI results,
but for the new data $\chi$MAID performs better.
\newline \indent At threshold and with the small momentum transfers in this 
experiment, chiral symmetry and its breaking should be a good framework for 
models. Therefore, our result is a clear challenge for theory. In view of the 
fundamental importance of the pion in hadron and nuclear physics a new effort 
is needed. 

\section*{Acknowledgments} \label{acknowledgments}
We are highly indebted to the MAMI accelerator group for the 
outstanding beam quality. We would like to thank S. Scherer and L. Tiator for 
helpful discussions. This work was supported in part by the Deutsche 
Forschungsgemeinschaft with the Collaborative Research Centres 443 and 1044 
and by the Croatian Science Foundation under project HRZZ 1680.

\bibliographystyle{elsarticle-num.bst}
\bibliography{pion_electroproduction_SOS}

\begin{thebibliography}{10}
\expandafter\ifx\csname url\endcsname\relax
  \def\url#1{\texttt{#1}}\fi
\expandafter\ifx\csname urlprefix\endcsname\relax\def\urlprefix{URL }\fi
\expandafter\ifx\csname href\endcsname\relax
  \def\href#1#2{#2} \def\path#1{#1}\fi

\bibitem{Nambu1962_1}
Y.~Nambu, D.~Luri\'{e}, Phys. Rev. 125 (1962) 1429.
\newblock \href {http://dx.doi.org/10.1103/PhysRev.125.1429}
  {\path{doi:10.1103/PhysRev.125.1429}}.

\bibitem{Nambu1962_2}
Y.~Nambu, E.~Shrauner, Phys. Rev. 128 (1962) 862.
\newblock \href {http://dx.doi.org/10.1103/PhysRev.128.862}
  {\path{doi:10.1103/PhysRev.128.862}}.

\bibitem{Amladi1979}
E.~Amaldi, S.~Fubini, G.~Furlan, Pion Electroproduction, Vol.~83 of Springer
  Tracts in Modern Physics, Springer, Berlin, 1979.
\newblock \href {http://dx.doi.org/10.1007/BFb0048208}
  {\path{doi:10.1007/BFb0048208}}.

\bibitem{Bernard1992}
V.~Bernard, N.~Kaiser, U.-G. Meissner, Phys. Rev. Lett. 69 (1992) 1877.
\newblock \href {http://dx.doi.org/10.1103/PhysRevLett.69.1877}
  {\path{doi:10.1103/PhysRevLett.69.1877}}.

\bibitem{Bernard1993}
V.~Bernard, N.~Kaiser, T.-S.~H. Lee, U.-G. Meissner, Phys. Rev. Lett. 70 (1993)
  387.
\newblock \href {http://dx.doi.org/10.1103/PhysRevLett.70.387}
  {\path{doi:10.1103/PhysRevLett.70.387}}.

\bibitem{Vainshtein1972}
A.~I. Vainshtein, V.~I. Zakharov, Nucl. Phys. B 36 (1972) 589.
\newblock \href {http://dx.doi.org/10.1016/0550-3213(72)90238-6}
  {\path{doi:10.1016/0550-3213(72)90238-6}}.

\bibitem{Scherer1991}
S.~Scherer, J.~H. Koch, Nucl. Phys. A 534 (1991) 461.
\newblock \href {http://dx.doi.org/10.1016/0375-9474(91)90456-G}
  {\path{doi:10.1016/0375-9474(91)90456-G}}.

\bibitem{Arndt_2006}
R.~A. Arndt, et~al., Phys. Rev. C 74 (2006) 045205.
\newblock \href {http://dx.doi.org/10.1103/PhysRevC.74.045205}
  {\path{doi:10.1103/PhysRevC.74.045205}}.

\bibitem{Ronchen_2013}
D.~R{\"o}nchen, et~al., Eur. Phys. J. A 49 (2013) 44.
\newblock \href {http://dx.doi.org/10.1140/epja/i2013-13044-5}
  {\path{doi:10.1140/epja/i2013-13044-5}}.

\bibitem{Drechsel_2007_47}
D.~Drechsel, S.~S. Kamalov, L.~Tiator, Eur. Phys. J. A 34 (2007) 69.
\newblock \href {http://dx.doi.org/10.1140/epja/i2007-10490-6}
  {\path{doi:10.1140/epja/i2007-10490-6}}.

\bibitem{Bloom1973}
E.~D. Bloom, et~al., Phys. Rev. Lett. 30 (1973) 1186.
\newblock \href {http://dx.doi.org/10.1103/PhysRevLett.30.1186}
  {\path{doi:10.1103/PhysRevLett.30.1186}}.

\bibitem{Esaulov1978}
A.~S. Esaulov, A.~M. Pilipenko, {\relax Yu}.~I. Titov, Nucl. Phys. B 136 (1978)
  511.
\newblock \href {http://dx.doi.org/10.1016/0550-3213(78)90273-0}
  {\path{doi:10.1016/0550-3213(78)90273-0}}.

\bibitem{Amladi1972}
E.~Amaldi, et~al., Phys. Lett. B 41 (1972) 216.
\newblock \href {http://dx.doi.org/10.1016/0370-2693(72)90465-0}
  {\path{doi:10.1016/0370-2693(72)90465-0}}.

\bibitem{Brauel1973}
P.~Brauel, et~al., Phys. Lett. B 45 (1973) 389.
\newblock \href {http://dx.doi.org/10.1016/0370-2693(73)90062-2}
  {\path{doi:10.1016/0370-2693(73)90062-2}}.

\bibitem{Joos1976}
P.~Joos, et~al., Phys. Lett. B 62 (1976) 230.
\newblock \href {http://dx.doi.org/10.1016/0370-2693(76)90514-1}
  {\path{doi:10.1016/0370-2693(76)90514-1}}.

\bibitem{delGuerra1975}
A.~del Guerra, et~al., Nucl. Phys. B 99 (1975) 253.
\newblock \href {http://dx.doi.org/10.1016/0550-3213(75)90004-8}
  {\path{doi:10.1016/0550-3213(75)90004-8}}.

\bibitem{delGuerra1976}
A.~del Guerra, et~al., Nucl. Phys. B 107 (1976) 65.
\newblock \href {http://dx.doi.org/10.1016/0550-3213(76)90191-7}
  {\path{doi:10.1016/0550-3213(76)90191-7}}.

\bibitem{Choi1993}
S.~Choi, et~al., Phys. Rev. Lett. 71 (1993) 3927.
\newblock \href {http://dx.doi.org/10.1103/PhysRevLett.71.3927}
  {\path{doi:10.1103/PhysRevLett.71.3927}}.

\bibitem{Park2012}
K.~Park, et~al., Phys. Rev. C 85 (2012) 035208.
\newblock \href {http://dx.doi.org/10.1103/PhysRevC.85.035208}
  {\path{doi:10.1103/PhysRevC.85.035208}}.

\bibitem{Blomqvist1996}
K.~I. Blomqvist, et~al., Z. Phys. A 353 (1996) 415.
\newblock \href {http://dx.doi.org/10.1007/BF01285153}
  {\path{doi:10.1007/BF01285153}}.

\bibitem{Liesenfeld1999}
A.~Liesenfeld, et~al., Phys. Lett. B 468 (1999) 20.
\newblock \href {http://dx.doi.org/10.1016/S0370-2693(99)01204-6}
  {\path{doi:10.1016/S0370-2693(99)01204-6}}.

\bibitem{Drechsel1992}
D.~Drechsel, L.~Tiator, J. Phys. G 18 (1992) 449.
\newblock \href {http://dx.doi.org/10.1088/0954-3899/18/3/004}
  {\path{doi:10.1088/0954-3899/18/3/004}}.

\bibitem{Rosenbluth1950}
M.~N. Rosenbluth, Phys. Rev. 79 (1950) 615.
\newblock \href {http://dx.doi.org/10.1103/PhysRev.79.615}
  {\path{doi:10.1103/PhysRev.79.615}}.

\bibitem{Blom1998}
K.~I. Blomqvist, et~al., Nucl. Instrum. Methods Phys. Res., Sect. A 403 (1998)
  263.
\newblock \href {http://dx.doi.org/10.1016/S0168-9002(97)01133-9}
  {\path{doi:10.1016/S0168-9002(97)01133-9}}.

\bibitem{Baumann2015}
D.~Baumann, et~al., (to be published).

\bibitem{Herminghaus1976}
H.~Herminghaus, et~al., Nucl. Instrum. Methods 138 (1976) 1.
\newblock \href {http://dx.doi.org/10.1016/0029-554X(76)90145-2}
  {\path{doi:10.1016/0029-554X(76)90145-2}}.

\bibitem{web_models_43}
\url{http://wwwkph.kph.uni-mainz.de/MAID/maid.html}.

\bibitem{Kamalov_44}
S.~S. Kamalov, et~al., Phys. Lett. B 522 (2001) 27.
\newblock \href {http://dx.doi.org/10.1016/S0370-2693(01)01241-2}
  {\path{doi:10.1016/S0370-2693(01)01241-2}}.

\bibitem{Kamalov_revC_45}
S.~S. Kamalov, et~al., Phys. Rev. C 64 (2001) 032201(R).
\newblock \href {http://dx.doi.org/10.1103/PhysRevC.64.032201}
  {\path{doi:10.1103/PhysRevC.64.032201}}.

\bibitem{Kamalov_revLett_46}
S.~S. Kamalov, S.~N. Yang, Phys. Rev. Lett. 83 (1999) 4494.
\newblock \href {http://dx.doi.org/10.1103/PhysRevLett.83.4494}
  {\path{doi:10.1103/PhysRevLett.83.4494}}.

\bibitem{Yang_2012}
S.~N. Yang, S.~S. Kamalov, L.~Tiator, AIP Conf. Proc. 1432 (2012) 293.
\newblock \href {http://dx.doi.org/10.1063/1.3701233}
  {\path{doi:10.1063/1.3701233}}.

\bibitem{Yang_2016}
S.~N. Yang, JPS Conf. Proc. 10 (2016) 042002.
\newblock \href {http://dx.doi.org/10.7566/JPSCP.10.042002}
  {\path{doi:10.7566/JPSCP.10.042002}}.

\bibitem{Hilt_2013_49}
M.~Hilt, B.~C. Lehnhart, S.~Scherer, L.~Tiator, Phys. Rev. C 88 (2013) 055207.
\newblock \href {http://dx.doi.org/10.1103/PhysRevC.88.055207}
  {\path{doi:10.1103/PhysRevC.88.055207}}.

\bibitem{Hilt_2016}
M.~Hilt, B.~C. Lehnhart, S.~Scherer, L.~Tiator, JPS Conf. Proc. 10 (2016)
  010007.
\newblock \href {http://dx.doi.org/10.7566/JPSCP.10.010007}
  {\path{doi:10.7566/JPSCP.10.010007}}.

\bibitem{James:2006zz}
F.~James, Statistical methods in experimental physics, 2nd Edition, World
  Scientific, 2006.

\end{thebibliography}

\end{document}